\documentclass[a4paper]{article}

\usepackage{INTERSPEECH2020}
\usepackage{multirow}
\usepackage{nccmath}

\usepackage{placeins}

\title{Speaker De-identification System using Autoencoders \\ and Adversarial Training}
\name{Fernando M. Espinoza-Cuadros$^{1,2}$, Juan M.~Perero-Codosero$^{1,2}$, Javier Antón-Martín$^{1,2}$, Luis A. Hernández-Gómez$^2$}

\address{
  $^1$Sigma Technologies S.L.U., Madrid, Spain\\
  $^2$GAPS Signal Processing Applications Group, Universidad Politécnica de Madrid, Madrid, Spain}
\email{\{fmespinoza,jmperero,janton\}@sigma-ai.com, luisalfonso.hernandez@upm.es}

\begin{document}

\maketitle
\begin{abstract}

The fast increase of web services and mobile apps, which collect personal data from users, increases the risk that their privacy may be severely compromised.\ In particular, the increasing variety of spoken language interfaces and voice assistants empowered by the vertiginous breakthroughs in Deep Learning are prompting important concerns in the European Union to preserve speech data privacy.\ For instance, an attacker can record speech from users and impersonate them to get access to systems requiring voice identification.\ Hacking speaker profiles from users is also possible by means of existing technology to extract speaker, linguistic (e.g.,\ dialect) and paralinguistic features (e.g.,\ age) from the speech signal. In order to mitigate these weaknesses, in this paper, we propose a speaker de-identification system based on adversarial training and autoencoders in order to suppress speaker, gender, and accent information from speech. Experimental results show that combining adversarial learning and autoencoders increase the equal error rate of a speaker verification system while preserving the intelligibility of the anonymized spoken content.

\end{abstract}

\section{Introduction}

Recent European privacy legislation, i.e., General Data Protection Regulation (GDPR), has limited some uses of the data in order to protect the personal information.\ According to the recent regulations, stored biometric data need to be unlinkable, irreversible, and renewable \cite{ISO/IEC2011}.\ This is the case of speech, which is considered personal information by itself.\ The main reason is that speech contains extra information apart from spoken contents.\ Furthermore, the emergent use of voice assistants has made that spoken commands are used by applications to carry out different actions.\ Sometimes it is necessary to collect some speech to improve and adapt the assistant's models to the user's speech.\ In this case, an attacker could have access to sensitive user’s data (e.g.,\ not only several utterances, but also some speaker profiles, such as age, gender.\ that can be easily obtained from these utterances). Thus, the objective is privacy preservation, suppressing critical speaker information from speech.

In order to preserve speaker privacy, some solutions have been proposed.\ Cryptography-based solutions involve a large complexity and computational overhead. Instead, anonymization is more flexible allowing also the removal of personally identifiable information within a speech signal.\ Since there is not a formal definition of anonymization (de-identification), VoicePrivacy initiative is defining metrics, protocols, and a benchmark on common datasets.\ Thus, privacy preservation solutions will be developed to anonymize the speech, but maintaining intelligibility and naturalness \cite{Tomashenko}.

The speaker de-identification task aims to suppress the speaker identity, which might be represented in the linguistic content of the speaker’s speech \cite{Parthasarathi2010,Parthasarathi2013} and spectral and excitation features of the speech signal \cite{Hashimoto2016, Jin2009}. Previous studies in speaker de-identification area are very limited. Most of them are based on voice transformation (VT) systems \cite{Jin2009,Pobar2014}, and phoneme recognition followed by speech synthesis from the phoneme sequence \cite{Fang2019}. In \cite{Pobar2014}, authors proposed an improved VT-based approach to enable the speaker to be de-identified by voice transformation from a pool of pre-trained VT models. In \cite{Fang2019}, authors proposed the de-identification of the real speaker by averaging a set of x-vectors from a pool of pre-trained x-vectors and select the most dissimilarity x-vector.\ This final x-vector along with the sequence of diphones and fundamental frequency (F0) synthesize the anonymized speech.

Domain-Adversarial training (DAT) \cite{Ganin2015} has been applied to improve Automatic Speech Recognition (ASR) performance by learning features invariant to various conditions, such as acoustic variabilities \cite{Serdyuk2016, Shinohara2016}, accented speech \cite{Sun2018}, and inter-speaker feature variability \cite{Adi2019,Meng2018,Tsuchiya2018}. Similarly, these techniques have been applied for speaker privacy protection. Speaker privacy protection in \cite{Srivastava2019} uses adversarial training to generate representations that perform well in ASR while hiding speaker identity. Along the same lines, in \cite{Saon2017} speaker-invariant training is carried out via reconstruction network in addition to the DNN acoustic model and trained jointly via adversarial training.\ Following the same approach, in this paper, we propose a speaker de-identification method based on the combination of adversarial training and autoencoders in order to generate speaker-invariant features as well as to other speaker characteristics (i.e., gender and accent).\ The rest of the paper is organized as follows.\ In Section 2, we describe the proposed speaker de-identification system based on x-vector anonymization. Section 3 explains the experimental setup under the VoicePrivacy 2020 Challenge \cite{Tomashenko}. Results are presented and discussed in Section 4. Finally, conclusions and future work are given in Section 5.

\section{Speaker de-identification system}

The speech signal contains different sources of variability. The speaker-dependent variability has been used to develop speaker characterization systems (e.g.,\ age \cite{Ghahremani2018}, gender \cite{Kotti2008}, pathologies \cite{botelho2020pathological, Perero-Codosero2019}, among others).\ In other cases, as in Automatic Speech Recognition (ASR), the speaker variability together with the acoustic environment variabilities (e.g.,\ noise, channels, etc.) are considered undesired sources of variability. This has led to the proposal of different techniques to remove the effect of speaker \cite{Sun2018,Adi2019,Meng2018,Tsuchiya2018} and the noise conditions in \cite{Serdyuk2016}, \cite{Shinohara2016} to improve the ASR accuracy.\ Similarly, in \cite{Saon2017}, the use of a reconstruction network an a DNN acoustic model is jointly optimized through adversarial multi-task learning to generate speaker-invariant features.

Following a similar approach, we propose a speaker de-identification method using DAT \cite{Ganin2015} and Autoencoders.\ Our method does not start from scratch, but it is based on the Baseline-1 anonymization system proposed in VoicePrivacy 2020 Challenge \cite{Tomashenko}, from now on referred as the baseline.\ It consists of three main parts:\ 1) feature extraction, 2) x-vector anonymization, and 3) speech synthesis.\ In this work we address part 2) of the baseline x-vector anonymization.\ In our approach, Adversarial Training and Autoencoders are proposed to remove information related to the speaker's characteristics in the anonymized x-vector while preserving an acceptable ASR performance.

\subsection{Autoencoder-Adversarial Network}

\begin{figure}[!ht]
\centering
\includegraphics[width=1\columnwidth]{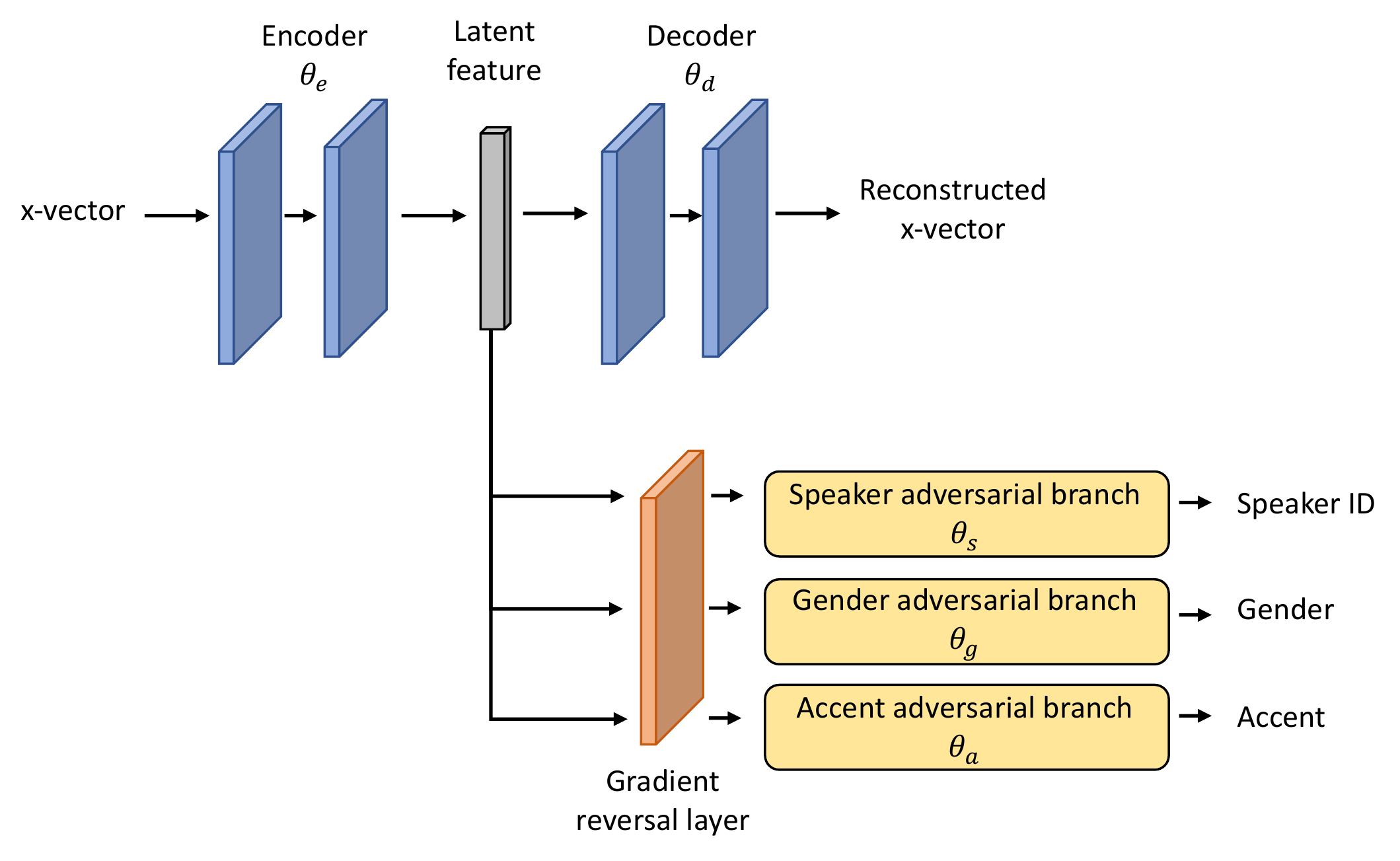}
\caption{Autoencoder-Adversarial Network (AAN) architecture.}
\label{fig:figure_1}
\end{figure}

Based on speaker de-identification methods using adversarial networks presented in \cite{Meng2018,Srivastava2019,Saon2017}, we propose a speaker-characteristics-invariant approach based on an Autoencoder-Adversarial Network (AAN). In the proposed ANN architecture (see Fig. \ref{fig:figure_1}), an encoder-decoder autoencoder branch tries to reconstruct the input x-vector while in adversarial branches we try to mitigate speaker characteristics, such as gender, accent and speaker identity.\ From this approach, we aim to hide the speaker identity when reconstructing the x-vector by means of the autoencoder but making the latent or encoded representation invariant to the domain of speaker characteristics by using a Domain Adversarial Neural Network (DANN) \cite{Ganin2015}.\ For this adversarial architecture, we are now given a training dataset denoted as $\left\{x_i,y_i,z_{gi},z_{ai},z_{si}\right\}_{N}^{i=1}$, where $x_i$ and $y_i$ are the original and reconstructed x-vector, respectively, and $z_{gi}$, $z_{ai}$, $z_{si}$ are the different domain classes as gender, accent, and speaker identity respectively of the i-th data point.\ We denote the $\theta _e$ and $\theta _d$ the parameters of the latent representation and decoder of the autoencoder respectively, and by $\theta _g$, $\theta _a$ and $\theta _s$ the parameters of the gender, accent and speaker identity of the adversarial branches respectively.\ The objective function for the autoencoder $L_{au}$ and adversarial branches $L_{z}$ are defined as

\begin{equation} 
\label{eq:1}
L_{au}(\theta _e, \theta _d)=-\sum_{i=i}^{N}\log P(y_i|x_i;\theta _e,\theta _d)
\end{equation}

\begin{equation}
\label{eq:2}
\resizebox{.9\hsize}{!}{$L_{z}(\theta _e,\theta _g, \theta _a, \theta _s)=-\sum _{k\in \left \{ g,a,s \right \}}\sum_{i}\log P(z_{si}|x_i;\theta _e,\theta _k)$}
\end{equation}
Thus, our model is trained by optimizing the following min-max objective:

\begin{equation} 
\label{eq:3}
\underset{\theta _e,\theta _d}{\min} \; \underset{\theta _e,\theta _g,\theta _a,\theta _s}{\max} \; \mathit{L_{au}(\theta _e,\theta _d)} - \lambda\mathit{L_{z}(\theta _e,\theta _g,\theta _a,\theta _s)}, 
\end{equation}

where $\lambda$ is a trade-off parameter between the autoencoder objective and the adversarial objectives, which goal is to remove the speaker characteristics via backpropagation by means the Gradient Reversal Layer (GRL) \cite{Ganin2015} in each adversarial branch. 

\begin{figure}[!ht]
\centering
\includegraphics[width=1\columnwidth]{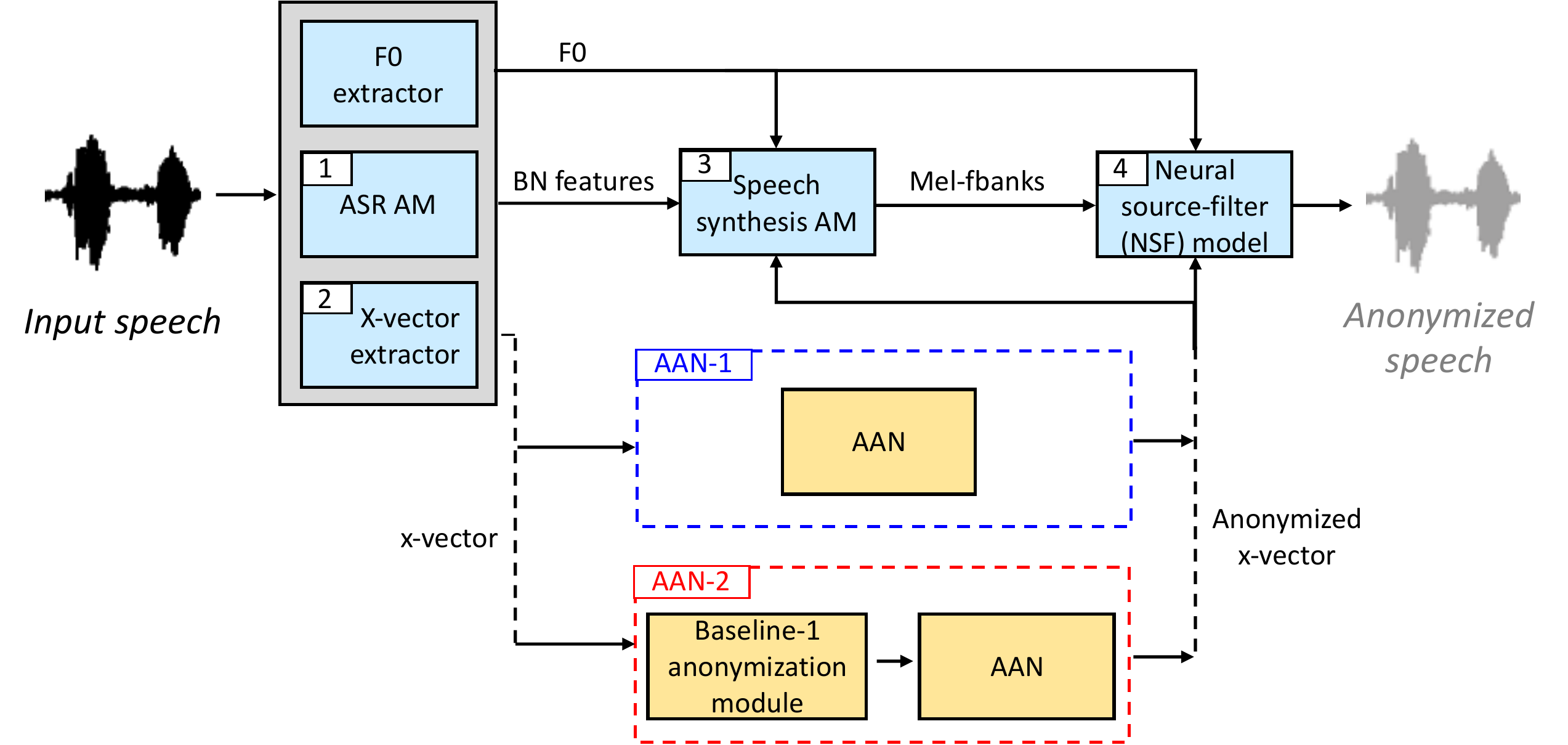}
\caption{X-vector anonymization approaches (i.e.\, AAN-1 and AAN-2) based on Autoencoder-Adversarial Networks. Adapted from \cite{Tomashenko}.}
\label{fig:figure_2}
\end{figure}

\subsection{X-vector anonymization approaches}

Based on the proposed framework, we evaluate two different approaches for x-vector anonymization.\ Both approaches use the x-vector extractor from the baseline system.\ In the first approach, the AAN described before, is used as x-vector anonymizer.\ That is, the x-vectors extracted from the baseline are used as input to the autoencoder that generates as output the pseudo-speaker x-vector, as shown in Fig. \ref{fig:figure_2} (AAN-1 label). In the second approach, we transform the pseudo-speaker x-vector generated by the baseline. As it can be seen in Fig. \ref{fig:figure_2} (AAN-2 label), the anonymized x-vector is used as input to the autoencoder that generates as output a new anonymized x-vector. 

\subsection{ANN Network architecture and training}

The enconder-decoder autoencoder model consists of 4 dense layers of size 512 with \textit{tanh} activation functions, trained using Mean Square Error (MSE) loss function. Each adversarial branch consists of a dense layer of size 128 and ReLU activation followed by a \textit{softmax} output layer, which output dimension corresponds to the number of classes on each adversarial feature. For the feature \textit{accent} the output dimension is 30, i.e., the number of accents in training dataset, for \textit{speaker} is 1251, i.e., speakers and for \textit{gender} is 2, i.e., female and male. Cross-entropy loss function was used on the adversarial training. 

For AAN training and testing we used VoxCeleb-1 \cite{Nagrani2017}, which contains approximately 330 hours of recordings from 1251 speakers. It also contains gender and accent information for each speaker. For AAN training, a closed-set speaker identification task was performed for the speaker adversarial branch. We assign 10 utterances per speaker to validation and test. The remaining utterances were used for training. For the rest of the adversarial branches, gender and accent labels were used. To select the optimal trade-off-parameter $\lambda$, several values were tested running the anonymization task of the VoicePrivacy 2020 Challenge. The best results were achieved for $\lambda=8$.

\section{Experimental setup}

\subsection{Dataset}

The proposed anonymization system, for both ANN-1 and ANN-2 approaches, was evaluated accordingly to VoicePrivacy 2020 Challenge using LibriSpeech \cite{Panayotov2015} and VCTK \cite{christophe2016cstr} datasets for both ASR (intelligibility) and Automatic Speaker Verification (ASV) (anonymization) evaluation tasks.\ A detailed description of both datasets used in the challenge can be found in \cite{Tomashenko}. 

\subsection{Evaluation system and metrics}

The anonymization system performance was evaluated by means of the assessment of the speaker verifiability and the ability of the anonymization system to preserve the intelligibility of the anonymized spoken content, which is carried out by pretrained ASV$_{eval}$ and ASR$_{eval}$ models provided by the VoicePrivacy Challenge.\ The metrics for both ASV and ASR tasks evaluation are Equal Error Rate (EER) and Log-likelihood-ratio cost function (Cllr), and Word error rate (WER), respectively.\ A detailed description of these metrics can be found in \cite{Tomashenko}. 
\begin{table}[b]
\centering
\caption{ASR results for \textbf{Baseline} for development and test data (o-original, a-anonymized speech).}
\resizebox{\columnwidth}{!}{
\begin{tabular}{|c|c|c|c||c||c|c|c|}
\hline
\multirow{2}{*}{\#} & \multirow{2}{*}{\textbf{Dev. set}} & \multicolumn{2}{c|}{\textbf{WER, \%}} & \multirow{2}{*}{\textbf{Data}} & \multirow{2}{*}{\textbf{Test set}} & \multicolumn{2}{c|}{\textbf{WER, \%}} \\ \cline{3-4} \cline{7-8}  &  & \textbf{$\text{LM}_{s}$} & \textbf{$\text{LM}_{l}$} &  &  & \textbf{$\text{LM}_{s}$} & \textbf{$\text{LM}_{l}$} \\ \hline\hline
1 & libri\_dev & 5.25 & 3.83 & o & libri\_test & 5.55 & 4.15\\ \hline 
2 & libri\_dev & 8.76 & 6.39 & a & libri\_test & 9.15 & 6.73\\ \hline 
\hline 3 & vctk\_dev & 14.00 & 10.79 & o & vctk\_test & 16.39 & 12.82\\ \hline 
4 & vctk\_dev & 18.92 & 15.38 & a & vctk\_test & 18.88 & 15.23\\ \hline 
\end{tabular}}
\label{table:table_1}
\end{table}

\section{Results and discussion}

In this section, we present the results for ASV (Tables \ref{table:table_5} and \ref{table:table_6}) and ASR (Tables \ref{table:table_2} and \ref{table:table_3}) tasks of the Challenge for both ANN-1 and ANN-2 anonymization approaches. We also compare our results to those from the baseline (Tables \ref{table:table_1} and \ref{table:table_4}). 

Overall, when compared to the baseline system, our results show that both ANN proposals increase speaker de-identification while providing similar intelligibility of the anonymized spoken content. However, the increase in speaker de-identification is only observed for both-sides anonymization condition (a-enroll, a-trial).\ That is when comparing to the baseline (Table \ref{table:table_4}), ANN results for the original-enroll and anonymized-trial condition (a-enroll, o-trial) in Tables \ref{table:table_5} and \ref{table:table_6} show, in the worst case, a decrease in speaker de-identification performance, in terms of EER, of approximately 9\% and 8\% for AAN-1 and AAN-2 respectively. In contrast, in the both-sides anonymization condition (a-enroll, a-trial), both approaches achieve better speaker de-identification results. For the best-case, we can observe an increase in performance over the baseline, in terms of EER, of approximately 9\% and 10\% for AAN-1 and AAN-2 respectively. Results in Table \ref{table:table_5} and \ref{table:table_6}, also show that for both best and worst- case scenarios, similar results are obtained in both ANN approaches.

Differences in performance between the ANNs approaches and the baseline for different anonymization conditions can be related to the performance of the proposed x-vector anonymization methods.\ For the o-enroll, a-trial condition, in the x-vector anonymization baseline, there is a high chance that the anonymized x-vector is very different from the original one as it corresponds to an average of farthest x-vectors from the original. Whereas in the AAN-1 approach, as the autoencoder aims to reconstruct the original x-vector while suppressing the speakers’ characteristics via adversarial training, there is a less probability that the reconstructed x-vector in the anonymized trials will be very far from the original.\ In the ANN-2 approach, we could expect that the anonymization performance should overcome the baseline due to the addition of variability to the anonymized x-vector. Nevertheless, that is not the case since the ANN training it is not optimized to reconstruct the anonymized x-vector.\ In contrast, in the both-sides anonymization condition (a-enroll, a-trial), we believe that the baseline system may have a high chance that the anonymized x-vectors in the enrollment and the trial fall in the same region since the anonymized x-vectors are selected from the farthest x-vectors from the original utterances on both sides, which belong to the same speaker. Thus, there is a chance that the anonymized x-vectors on both sides might be closed to each other.\ Differently from that, in our approach, the use of adversarial training for AAN introduces and additional variability to the reconstructed x-vector that can lead to the observed increase in speaker de-identification.

\begin{table}[h!]
\centering
\caption{ASR results for \textbf{AAN-1} for development and test data (a-anonymized speech).}
\resizebox{\columnwidth}{!}{
\begin{tabular}{|c|c|c|c||c||c|c|c|}
\hline
\multirow{2}{*}{\#} & \multirow{2}{*}{\textbf{Dev. set}} & \multicolumn{2}{c|}{\textbf{WER, \%}} & \multirow{2}{*}{\textbf{Data}} & \multirow{2}{*}{\textbf{Test set}} & \multicolumn{2}{c|}{\textbf{WER, \%}} \\ \cline{3-4} \cline{7-8}  &  & \textbf{$\text{LM}_{s}$} & \textbf{$\text{LM}_{l}$} &  &  & \textbf{$\text{LM}_{s}$} & \textbf{$\text{LM}_{l}$} \\ \hline \hline 
1 & libri\_dev & 9.22 & 6.75 & a & libri\_test & 9.24 & 6.74\\ \hline 
 \hline 
2 & vctk\_dev & 18.67 & 15.20 & a & vctk\_test & 19.09 & 15.16\\ \hline 
\end{tabular}
}
\label{table:table_2}
\end{table}

\begin{table}[h!]
\centering
\caption{ASR results for \textbf{AAN-2} for development and test data (a-anonymized speech).}
\resizebox{\columnwidth}{!}{
\begin{tabular}{|c|c|c|c||c||c|c|c|}
\hline
\multirow{2}{*}{\#} & \multirow{2}{*}{\textbf{Dev. set}} & \multicolumn{2}{c|}{\textbf{WER, \%}} & \multirow{2}{*}{\textbf{Data}} & \multirow{2}{*}{\textbf{Test set}} & \multicolumn{2}{c|}{\textbf{WER, \%}} \\ \cline{3-4} \cline{7-8}  &  & \textbf{$\text{LM}_{s}$} & \textbf{$\text{LM}_{l}$} &  &  & \textbf{$\text{LM}_{s}$} & \textbf{$\text{LM}_{l}$} \\ \hline \hline 
1 & libri\_dev & 9.28 & 6.76 & a & libri\_test & 9.37 & 6.85\\ \hline 
 \hline 
2 & vctk\_dev & 18.69 & 15.25 & a & vctk\_test & 19.04 & 15.21\\ \hline 
\end{tabular}
}
\label{table:table_3}
\end{table}

Finally, as stated before, results for ANN-1 and ANN-2 in the ASR task (Tables \ref{table:table_2} and \ref{table:table_3}) show a performance in intelligibility of the anonymized spoken content similar to that of the baseline system (Table \ref{table:table_1}).

\section{Conclusions and future work}

In this work, we present two methods for x-vector anonymization.\ These methods are integrated and evaluated under the Baseline-1 anonymization system proposed in VoicePrivacy 2020 Challenge.\ Both methods rely on an Autoencoder-Adversarial Network that tries to reconstruct x-vectors but intending to alleviate, through adversarial branches, the information of speaker characteristics in order to hide the speaker identity to a greater extent. Our experimental results show that though similar results to the baseline were achieved, when testing on both sides anonymization condition (i.e., training and testing with anonymized speech) our system outperforms the baseline.\ Those results foster to keep researching in adversarial training techniques, as well as the use of generative models to generate speaker-invariant features. Our future research will address the application of this framework to both the encoded speech content and the prosodic features looking for a better anonymization of the speech waveform suppressing the speaker information but preserving the spoken content.

\begin{table*}[]
\centering
\caption{ASV results for \textbf{Baseline} for development and test data (o-original, a-anonymized speech; \textbf{Gen} denotes speaker gender: \textbf{f}-female, \textbf{m}-male).}
\normalsize
\begin{tabular}{|c|c|c|c|c||c|c|c||c|c|c|c|}
\hline
\# & \textbf{Dev. set} & \textbf{EER, \%} & \textbf{$\text{C}_{llr}^{min}$} & \textbf{$\text{C}_{llr}$} & \textbf{Enroll} & \textbf{Trial} & \textbf{Gen} & \textbf{Test set} & \textbf{EER, \%} & \textbf{$\text{C}_{llr}^{min}$} & \textbf{$\text{C}_{llr}$}\\ \hline\hline
1 & libri\_dev & 8.665 & 0.304 & 42.857 & o & o & f & libri\_test & 7.664 & 0.183 & 26.793\\ \hline
2 & libri\_dev & 50.140 & 0.996 & 144.112 & o & a & f & libri\_test & 47.260 & 0.995 & 151.822\\ \hline
3 & libri\_dev & 36.790 & 0.894 & 16.345 & a & a & f & libri\_test & 32.120 & 0.839 & 16.270\\ \hline
\hline 4 & libri\_dev & 1.242 & 0.034 & 14.250 & o & o & m & libri\_test & 1.114 & 0.041 & 15.303\\ \hline
5 & libri\_dev & 57.760 & 0.999 & 168.988 & o & a & m & libri\_test & 52.120 & 0.999 & 166.658\\ \hline
6 & libri\_dev & 34.160 & 0.867 & 24.715 & a & a & m & libri\_test & 36.750 & 0.903 & 33.928\\ \hline
\hline 7 & vctk\_dev\_com & 2.616 & 0.088 & 0.868 & o & o & f & vctk\_test\_com & 2.890 & 0.091 & 0.866\\ \hline
8 & vctk\_dev\_com & 49.710 & 0.995 & 172.049 & o & a & f & vctk\_test\_com & 48.270 & 0.994 & 162.531\\ \hline
9 & vctk\_dev\_com & 27.910 & 0.741 & 7.205 & a & a & f & vctk\_test\_com & 31.210 & 0.830 & 9.015\\ \hline
\hline 10 & vctk\_dev\_com & 1.425 & 0.050 & 1.559 & o & o & m & vctk\_test\_com & 1.130 & 0.036 & 1.041\\ \hline
11 & vctk\_dev\_com & 54.990 & 0.999 & 192.924 & o & a & m & vctk\_test\_com & 53.390 & 1.000 & 190.136\\ \hline
12 & vctk\_dev\_com & 33.330 & 0.840 & 23.891 & a & a & m & vctk\_test\_com & 31.070 & 0.835 & 21.680\\ \hline
\hline 13 & vctk\_dev\_dif & 2.864 & 0.100 & 1.134 & o & o & f & vctk\_test\_dif & 4.887 & 0.169 & 1.495\\ \hline
14 & vctk\_dev\_dif & 49.970 & 0.989 & 166.027 & o & a & f & vctk\_test\_dif & 48.050 & 0.998 & 146.929\\ \hline
15 & vctk\_dev\_dif & 26.110 & 0.760 & 8.414 & a & a & f & vctk\_test\_dif & 31.740 & 0.847 & 11.527\\ \hline
\hline 16 & vctk\_dev\_dif & 1.439 & 0.052 & 1.158 & o & o & m & vctk\_test\_dif & 2.067 & 0.072 & 1.817\\ \hline
17 & vctk\_dev\_dif & 53.950 & 1.000 & 167.511 & o & a & m & vctk\_test\_dif & 53.850 & 1.000 & 167.824\\ \hline
18 & vctk\_dev\_dif & 30.920 & 0.839 & 23.797 & a & a & m & vctk\_test\_dif & 30.940 & 0.834 & 23.842\\ \hline
\end{tabular}
\label{table:table_4}
\end{table*}

\begin{table*}[]
\centering
\caption{ASV results for \textbf{AAN-1} for development and test data (o-original, a-anonymized speech; \textbf{Gen} denotes speaker gender: \textbf{f}-female, \textbf{m}-male).}
\normalsize
\begin{tabular}{|c|c|c|c|c||c|c|c||c|c|c|c|}
\hline
\# & \textbf{Dev. set} & \textbf{EER, \%} & \textbf{$\text{C}_{llr}^{min}$} & \textbf{$\text{C}_{llr}$} & \textbf{Enroll} & \textbf{Trial} & \textbf{Gen} & \textbf{Test set} & \textbf{EER, \%} & \textbf{$\text{C}_{llr}^{min}$} & \textbf{$\text{C}_{llr}$}\\ \hline\hline
1 & libri\_dev & 44.320 & 0.974 & 171.463 & o & a & f & libri\_test & 43.980 & 0.972 & 168.557\\ \hline
2 & libri\_dev & 39.630 & 0.921 & 22.336 & a & a & f & libri\_test & 34.850 & 0.886 & 27.144\\ \hline
 \hline
3 & libri\_dev & 49.840 & 0.989 & 153.223 & o & a & m & libri\_test & 45.430 & 0.980 & 155.451\\ \hline
4 & libri\_dev & 43.480 & 0.964 & 36.897 & a & a & m & libri\_test & 46.100 & 0.979 & 47.663\\ \hline
 \hline
5 & vctk\_dev\_com & 50.580 & 0.976 & 183.167 & o & a & f & vctk\_test\_com & 47.110 & 0.982 & 171.678\\ \hline
6 & vctk\_dev\_com & 29.650 & 0.802 & 14.289 & a & a & f & vctk\_test\_com & 37.570 & 0.913 & 17.304\\ \hline
 \hline
7 & vctk\_dev\_com & 47.860 & 0.985 & 171.920 & o & a & m & vctk\_test\_com & 44.920 & 0.984 & 172.326\\ \hline
8 & vctk\_dev\_com & 38.180 & 0.921 & 30.378 & a & a & m & vctk\_test\_com & 37.290 & 0.927 & 30.642\\ \hline
 \hline
9 & vctk\_dev\_dif & 49.860 & 0.957 & 177.802 & o & a & f & vctk\_test\_dif & 48.350 & 0.997 & 155.964\\ \hline
10 & vctk\_dev\_dif & 30.430 & 0.828 & 14.852 & a & a & f & vctk\_test\_dif & 34.000 & 0.881 & 21.306\\ \hline
 \hline
11 & vctk\_dev\_dif & 44.760 & 0.988 & 151.207 & o & a & m & vctk\_test\_dif & 48.160 & 0.996 & 157.427\\ \hline
12 & vctk\_dev\_dif & 33.800 & 0.882 & 30.406 & a & a & m & vctk\_test\_dif & 39.040 & 0.947 & 33.550\\ \hline
\end{tabular}
\label{table:table_5}
\end{table*}

\begin{table*}[]
\centering
\caption{ASV results for \textbf{AAN-2} for development and test data (o-original, a-anonymized speech; \textbf{Gen} denotes speaker gender: \textbf{f}-female, \textbf{m}-male).}
\normalsize
\begin{tabular}{|c|c|c|c|c||c|c|c||c|c|c|c|}
\hline
\# & \textbf{Dev. set} & \textbf{EER, \%} & \textbf{$\text{C}_{llr}^{min}$} & \textbf{$\text{C}_{llr}$} & \textbf{Enroll} & \textbf{Trial} & \textbf{Gen} & \textbf{Test set} & \textbf{EER, \%} & \textbf{$\text{C}_{llr}^{min}$} & \textbf{$\text{C}_{llr}$}\\ \hline
\hline
1 & libri\_dev & 45.880 & 0.981 & 171.212 & o & a & f & libri\_test & 44.890 & 0.980 & 166.823\\ \hline
2 & libri\_dev & 39.630 & 0.924 & 23.006 & a & a & f & libri\_test & 35.770 & 0.898 & 27.617\\ \hline
 \hline
3 & libri\_dev & 50.000 & 0.992 & 153.313 & o & a & m & libri\_test & 45.880 & 0.985 & 155.770\\ \hline
4 & libri\_dev & 43.940 & 0.968 & 38.383 & a & a & m & libri\_test & 46.770 & 0.979 & 48.320\\ \hline
 \hline
5 & vctk\_dev\_com & 50.870 & 0.981 & 183.799 & o & a & f & vctk\_test\_com & 46.820 & 0.983 & 172.018\\ \hline
6 & vctk\_dev\_com & 29.070 & 0.815 & 15.106 & a & a & f & vctk\_test\_com & 39.020 & 0.917 & 17.719\\ \hline
 \hline
7 & vctk\_dev\_com & 47.580 & 0.986 & 172.169 & o & a & m & vctk\_test\_com & 45.480 & 0.987 & 172.479\\ \hline
8 & vctk\_dev\_com & 39.320 & 0.936 & 31.763 & a & a & m & vctk\_test\_com & 38.700 & 0.943 & 32.110\\ \hline
 \hline
9 & vctk\_dev\_dif & 50.480 & 0.963 & 179.226 & o & a & f & vctk\_test\_dif & 49.280 & 0.996 & 156.662\\ \hline
10 & vctk\_dev\_dif & 31.050 & 0.833 & 15.638 & a & a & f & vctk\_test\_dif & 34.830 & 0.895 & 21.842\\ \hline
 \hline
11 & vctk\_dev\_dif & 45.310 & 0.991 & 151.928 & o & a & m & vctk\_test\_dif & 48.390 & 0.996 & 157.517\\ \hline
12 & vctk\_dev\_dif & 34.690 & 0.897 & 31.013 & a & a & m & vctk\_test\_dif & 39.610 & 0.955 & 33.957\\ \hline
\end{tabular}

\label{table:table_6}
\end{table*}

\clearpage
\bibliographystyle{IEEEtran}

\bibliography{VoicePrivacy_references}

\end{document}